\newcommand{\diracslash}[1]{#1\llap{/\kern2pt}}

\newcommand{\be}{\begin{equation}}
\newcommand{\ee}{\end{equation}}
\newcommand{\bea}{\begin{eqnarray}}
\newcommand{\eea}{\end{eqnarray}}
\newcommand{\ba}[1]{\begin{array}{#1}}
\newcommand{\ea}{\end{array}}

\documentclass[prd,aps,floats,nofootinbib,tightenlines,showpacs]{revtex4}
\usepackage{epsfig,graphicx,pstricks}
\usepackage{psfrag}
\usepackage{color}
\usepackage{amsmath}
\usepackage{amsfonts}
\usepackage{amssymb}
\usepackage{textcomp}
\usepackage{multirow}
\usepackage{natbib}
\usepackage{subfigure}
\begin{document}

\title { Fluctuations of conserved charges with finite size PNJL model }
\author{Paramita Deb }
\email{paramita.deb83@gmail.com}
\affiliation{Department of Physics, Indian Institute of Technology Bombay, Powai, Mumbai- 400076, India}
\author{Raghava Varma}
\email{varma@phy.iitb.ac.in}
\affiliation{Department of Physics, Indian Institute of Technology Bombay Powai, Mumbai-400076, India}

\date{\today} 

\def\be{\begin{equation}}
\def\ee{\end{equation}}
\def\bearr{\begin{eqnarray}}
\def\eearr{\end{eqnarray}}
\def\zbf#1{{\bf {#1}}}
\def\bfm#1{\mbox{\boldmath $#1$}}
\def\hf{\frac{1}{2}}
\def\sl{\hspace{-0.15cm}/}
\def\omit#1{_{\!\rlap{$\scriptscriptstyle \backslash$}
{\scriptscriptstyle #1}}}
\def\vec#1{\mathchoice
        {\mbox{\boldmath $#1$}}
        {\mbox{\boldmath $#1$}}
        {\mbox{\boldmath $\scriptstyle #1$}}
        {\mbox{\boldmath $\scriptscriptstyle #1$}}
}

\begin{abstract}
 Fluctuations of baryon, charge and strangeness have been investigated in Polyakov loop enhanced Nambu--Jona-Lasinio model. Multiple reflection expansion method has been incorporated to include the surface and curvature energy along with the system volume. The results of different fluctuations are then compared with the available experimental data from the heavy ion collision.
\end{abstract}


\maketitle

\section{Introduction}

The strongly interacting matter have a rich phase structure at finite 
temperature and density. Few microseconds after the Big Bang our Universe might be in a quark and gluon
 plasma phase, which then expanded and cooled to form our present hadronic state.
 One of the fundamental goals of the heavy ion collision experiments at Relativistic Heavy Ion Collider (RHIC) at BNL and Super Proton Synchroton (SPS) at CERN, is to map the QCD phase diagram and estimate the critical
  end point (CEP), where the first order phase transition from
  hadronic state to quark gluon plasma (QGP) phase becomes continuous
  {\cite{STAR, STAR1}}. Also choosing the correct experimental observables that will help to locate the critical point is equally important. In the heavy ion
   collision experiments, formation of QGP is followed by expansion and
   attainment of freeze out characterized by cessation of all chemical
   and kinetic interactions. The detected particles at the freeze out
   condition can lead to the location of the freeze out point. Thus to
   locate the critical point, experiments are being conducted to
   bring the freeze out point close to the critical point by varying
   the collision center of mass energy $\sqrt s$ \cite{STAR2_muB}.
   Therefore, there is a need to select the suitable experimental
   observables such as fluctuations of the conserved quantities that are
   sensitive to the proximity of the freeze-out point. The fluctuations of
   an experimental observables are defined as the variance and higher
   non-Gaussian moments of the event-by-event distribution of the
   observable at each event in ensemble of many events. At the critical
   point fluctuations and therefore, long range correlations at all length
   scales arises; \text{with a maximal value, $\xi \approx 1.5 - 3 fm$},
   due to the finite size and time slowing down effects in the heavy ion
   collisions \cite {rajagopal2000}. Further more, the magnitude of
   fluctuations of the conserved quantities (net-baryon $(\Delta{B})$,
   net-strangeness $(\Delta{S})$ and net-charge $(\Delta{Q})$) and the
   correlation length ($\xi$) diverge at the critical point. Therefore, the
   non-monotonic behaviour of these fluctuations of the higher order
   moments of these conserved quantities could be the signature of the
   critical point \cite{rajagopal1998}.

The finite volume of the QGP matter formed in the heavy ion collision experiment depends on the center of mass energy and the colliding nuclei. To estimate the finite volume produce in the collisions, several efforts have been made by centrality measurements of HBT radii {\cite {adamova}}. These measurements suggest that with the centrality the volume increases during freeze out and it is estimated to be $2000 fm^3$ to $3000 fm^3$. Theoretically the effects of finite volume have been addressed by many models such as non-interacting bag model {\cite {elze}}, chiral perturbation theory {\cite {luscher,gasser}}, Nambu--Jona-Lasinio (NJL) model {\cite {nambu,kiriyama,shao}}, linear sigma model {\cite {braun,braun1}} and  by the first principle study of pure gluon theory on space time lattices {\cite {gopie,bazavov}}. Specifically, in a $1+1$ dimensional NJL model the finite size effect of a dense baryonic matter has been described by the induction of a charged pion condensation phenomena. Recently, this has been extended to Polyakov loop Nambu--Jona-Lasinio (PNJL) model where it was observed that as the volume decreases, critical temperature for the crossover transition decreases. For lower volumes, CEP is shifted to a domain with  higher chemical potential $(\mu)$ and lower temperature $(T)$ {\cite {deb1,abhijit}. It is quite evident that broadly both the Lattice calculations and QCD-based models indicate that the fluctuation of strongly interacting matter at zero density show significant volume dependence which might be relevant to study the formation of fireball in heavy-ion collisions.

In the earlier version of PNJL model, we have used finite volume contribution a boundary condition in the lower limit of the integral of 
thermodynamic potential $p_{min}=\pi/R=\lambda$, where $R$ is the lateral size of the finite volume system {\cite{deb1}}. But several simplifications have been made in this type of finite volume system. We have neglected the surface and curvature effects, the infinte sum has been considered as an
integration over a continuous variation of momentum albeit with the lower
cut-off. Also any modifications to the mean-field parameters due to
finite size effects have not been considered. In order to improve our 
finite size study, we have considered the surface and curvature effects 
that bound the finite volume of PNJL system through multiple reflection expansion (MRE) method {\cite {bloch,madsen,kiriyama,grunfeld}}. Compared to previous finite volume
calculations, MRE describes a sphere with proper boundary conditions rather
than a cube, which is more  natural to study the problem of QGP  fireball in heavy ion collision.

Further, both the Lattice QCD results {\cite{boyd,engels,fodor,allton,forcrand,aoki,megias}} and the QCD inspired models {\cite{fukushima,ratti,pisarski,fukushima1,hansen,ciminale,ghosh,deb,deb1,osipov,kashiwa,schaefer, mustafa}} show that the net conserved quantum numbers ($B$, $Q$ and $S$) are related to the conserved number susceptibilities ($\chi_x= \langle(\delta N_x)^2\rangle/VT $ where $x$ can be either $B$, $S$ or $Q$ 
and $V$ is the volume). Close to the critical point, models also predict that the distributions of the conserved quantum numbers to be non-Gaussian and susceptibilities to diverge. Higher non-Gaussian moments of these conserved quantities such as skewness, $S$ ($\sim \xi^{4.5}$), and kurtosis, ($\sim \xi^{7.0}$) are related to the susceptibilities and much more sensitive to the correlation length which can be very useful in the search of CEP. As these higher order moments are volume dependent, moment products, such as, $s\sigma = (\chi_x^3/\chi_x^2)$ and $\kappa\sigma^2 = \chi_x^4 / \chi_x^2$ can be constructed to cancel out the system size dependency. Similarly, fluctuations have been also computed with respect to the quark chemical potential $(\mu)$ in the  Polyakov loop coupled quark-meson (PQM) model {\cite{schaefer}} and its renormalized group improved version, 2 flavor PNJL model with three-momentum cutoff regularization {\cite{ejiri,roessner}}. Fluctuations and the correlations of conserved charges have also been studied in higher flavor PNJL model \cite {deb,deb1, abhijit, upadhaya} with or without finite volume effects as well as simplistic lattice QCD  {\cite{cheng}}. Recently, a realistic continuum limit calculation \cite {bazavov1,bazavov2,borsanyi} for the lattice QCD data has been performed and the 3 flavor PNJL model parameters have been reconsidered \cite{saha}. However, in order to understand the QCD phase diagram and find the critical end point one need to explore the finite density region. Current work will emphasis on the 3 flavor finite size finite density PNJL model with MRE formalism to study the susceptibilities of different conserved charges and compare them to both the recent experimental finding.
 
In the context of the above discussions, we organize the present work as follows. We describe the thermodynamic formulation of the 3-flavor finite size PNJL model with MRE formalism. Subsequently, the method to calculate the correlation of conserved charges in PNJL model has been elaborated. Then the variation of skewness $(s)$, kurtosis $(\kappa)$ of fluctuations of  different conserved charges $(\chi)$ and higher moments with collision energy has  been determined. Finally, we have discussed the possible location of critical end point (CEP) by studying the QCD phase diagram with temperature and finite baryon density.

\section{The PNJL model}
We shall consider the 2+1 flavor PNJL model with six quark and eight quark interactions. In the PNJL model the gluon dynamics is described by the chiral point couplings between quarks (present in the NJL part) and a background gauge field representing Polyakov Loop dynamics. The Polyakov line is represented as,
\begin {equation}
  L(\bar x)={\cal P} {\rm exp}[i {\int_0}^\beta
d\tau A_4{({\bar x},\tau)}]
\end {equation}
where $A_4=iA_0$ is the temporal component of Eucledian gauge field
$(\bar A,A_4)$, $\beta=\frac {1}{T} $, and $\cal P$ denotes path
ordering. $L(\bar x)$ transforms as a field with charge one under
global Z(3) symmetry. The Polyakov loop is then given by 
$\Phi = (Tr_c L)/N_c$, and its conjugate by,
${\bar \Phi} = (Tr_c L^\dagger)/N_c$. The gluon dynamics can be
described as an effective theory of the Polyakov loops. Consequently,
the Polyakov loop potential can be expressed as,
\begin{equation}
\frac {{\cal {U^\prime}}(\Phi[A],\bar \Phi[A],T)} {T^4}= 
\frac  {{\cal U}(\Phi[A],\bar \Phi[A],T)}{ {T^4}}-
                                     \kappa \ln(J[\Phi,{\bar \Phi}])
\label {uprime}
\end{equation}
where $\cal {U(\phi)}$ is a Landau-Ginzburg type potential commensurate
with the Z(3) global symmetry. Here we choose a form given in
\cite{ratti},
\begin{equation}
\frac  {{\cal U}(\Phi, \bar \Phi, T)}{  {T^4}}=-\frac {{b_2}(T)}{ 2}
                 {\bar \Phi}\Phi-\frac {b_3}{ 6}(\Phi^3 + \bar \Phi^3)
                 +\frac {b_4}{  4}{(\bar\Phi \Phi)}^2,
\end{equation}
where
\begin {eqnarray}
     {b_2}(T)=a_0+{a_1}exp(-a2{\frac {T}{T_0}}){\frac {T_0}{T}}
\end {eqnarray}
$b_3$ and $b_4$ being constants. The second term in eqn.(\ref {uprime})
is the Vandermonde term which replicates the effect of SU(3) Haar
measure and is given by,
\begin {equation}
J[\Phi, {\bar \Phi}]=(27/24{\pi^2})\left[1-6\Phi {\bar \Phi}+\nonumber\\
4(\Phi^3+{\bar \Phi}^3)-3{(\Phi {\bar \Phi})}^2\right]
\end{equation}
The corresponding parameters were earlier obtained in the above
mentioned literature by choosing suitable values by fitting a few
physical quantities as function of temperature obtained in LQCD
computations. The set of values chosen here are listed in the 
table \ref{table1} \cite{saha}. 
\vskip 0.1in
\begin{table}[htb]
\begin{center}
\begin{tabular}{|c|c|c|c|c|c|c|c|c|c|c|c|}
\hline
Interaction & $ T_0 (MeV) $ & $ a_0 $ & $ a_1 $ & $ a_2 $ & $ b_3 $ &$
b_4$ & $  \kappa $ \\ 

\hline
6-quark &$ 175 $&$ 6.75 $&$ -9.0 $&$ 0.25 $&$ 0.805 $&$7.555 $&$ 0.1 $ \\

\hline

\end{tabular}
\caption{Parameters for the Polyakov loop potential of the model.}  
\label{table1}
\end{center}
\end{table}
\vskip 0.1in

 For the quarks we shall use the usual form of the NJL model except
for the substitution of a covariant derivative containing a background
temporal gauge field. Thus the 2+1 flavor the Lagrangian may be written as,
\begin{equation}
\begin{split}
   {\cal L} = {\sum_{f=u,d,s}}{\bar\psi_f}\gamma_\mu iD^\mu
             {\psi_f}&-\sum_f m_{f}{\bar\psi_f}{\psi_f}
              +\sum_f \mu_f \gamma_0{\bar \psi_f}{\psi_f}
       +{{g_S}\over{2}} {\sum_{a=0,\ldots,8}}[({\bar\psi} \lambda^a
        {\psi})^2+
            ({\bar\psi} i\gamma_5\lambda^a {\psi})^2] \nonumber\\
       &-{g_D} [det{\bar\psi_f}{P_L}{\psi_{f^\prime}}+det{\bar\psi_f}
            {P_R}{\psi_{f^\prime}}]\nonumber\\
                 &-{\cal {U^\prime}}(\Phi[A],\bar \Phi[A],T)
\end{split}
\end{equation}
where $f$ denotes the flavors $u$, $d$ or $s$ respectively.
The matrices $P_{L,R}=(1\pm \gamma_5)/2$ are respectively the
left-handed and right-handed chiral projectors, and the other terms
have their usual meaning, described in details in
Refs.~\cite{ghosh,deb,deb1}. This NJL part of the theory
is analogous to the BCS theory of superconductor, where the
pairing of two electrons leads to the condensation causing a gap in
the energy spectrum. Similarly in the chiral limit, NJL model exhibits
dynamical breaking of ${SU(N_f)}_L \times {SU(N_f)_R}$ symmetry to
$SU(N_f)_V$ symmetry ($N_f$ being the number of flavors). As a result 
the composite operators ${\bar \psi_f}\psi_f$ generate nonzero vacuum
expectation values. The quark condensate is given as,
\begin {equation}
 \langle{\bar \psi_f}{\psi_f}\rangle= 
-i{N_c}{{{\cal L}t}_{y\rightarrow x^+}}(tr {S_f}(x-y)),
\end {equation}
where trace is over color and spin states. The self-consistent gap 
equation for the constituent quark masses are,
\begin {equation}
  M_f =m_f-g_S \sigma_f+g_D \sigma_{f+1}\sigma_{f+2}  
\end {equation}
where $\sigma_f=\langle{\bar \psi_f} \psi_f\rangle$ denotes chiral 
condensate of the quark with flavor $f$. Here if we consider
$\sigma_f=\sigma_u$, then $\sigma_{f+1}=\sigma_d$ and
$\sigma_{f+2}=\sigma_s$, 
The expression for $\sigma_f$ at zero
temperature ($T=0$) and chemical potential ($\mu_f=0$) may be
written as \cite{deb},
\begin {equation}
 \sigma_f=-\frac {3{M_f}}{ {\pi}^2} {{\int}^\Lambda}\frac {p^2}{
           \sqrt {p^2+{M_f}^2}}dp,
\end {equation}
$\Lambda$ being the three-momentum cut-off. This cut-off have been used
to regulate the model because it contains couplings with finite dimensions
which leads to the model to be non-renormalizable.

Due to the dynamical breaking of chiral symmetry, $N_f^2 - 1$ Goldstone
bosons appear. These Goldstone bosons are the pions and kaons whose masses, 
decay widths from experimental observations are utilized to fix the NJL model
parameters. The parameter values have been listed in table \ref{table2}.
Here we consider the $\Phi$, $\bar \Phi$ and $\sigma_f$ fields in the
mean field approximation (MFA) where the mean field are obtained by
simultaneously solving the respective saddle point equations.

\begin{table}[htb]
\begin{center}
\begin{tabular}{|c|c|c|c|c|c|c|c|c|c|c|}
\hline
Model & $ m_u (MeV) $ & $ m_s (MeV)$ & $ \Lambda (MeV) $ & $ g_S \Lambda^2 $ & $ g_D \Lambda^5 $  \\ 

\hline
With 6-quark &$ 5.5 $&$ 134.76 $&$ 631 $&$ 3.67 $&$ 9.33 $ \\
\hline

\end{tabular}
\caption{Parameters of the Fermionic part of the model.}  
\label{table2}
\end{center}
\end{table}

 Now that the PNJL model is described for infinite volumes we discuss
how we implement the finite size constraints through MRE formalism.
\vskip 0.2in
 
\section{Multiple Reflection Expansion}
Multiple reflection expansion technique was first formulated by \cite{bloch}
in order to find out the distribution of eigenvalues of the equation 
$\bigtriangleup \phi + E \phi=0$ for an arbitrary volume and for  
boundary condition
$\partial \phi/ \partial n = \kappa \phi$ on the surface sufficiently smooth.
This multiple reflection expansion formalism (MRE) takes 
into account the modification in the density of states resulting the
system to be restricted in a finite domain \cite{bloch,madsen,kiriyama,
kiriyama1,grunfeld,zhao}. For the case of a finite spherical droplet the density 
of states reads,
\begin{equation}
\frac {dN_i} {dp}= 6[\frac {p^2V} {2\pi^2}+ f_S (\frac {m_i} {p}) kS 
+f_C (\frac {m_i} {p}) C + \ldots]
                 =\frac { p^2 \rho_{MRE}} {2\pi^2}
\label{dndk} 
\end{equation}
where area $S=\oint dS =4 \pi R^2$ and curvature $C = \oint (\frac {1}{R_1} +
 \frac {1} {R_2}) dS = 8 \pi R$ for a sphere. Curvature radii are denoted by
$R_1$ and $R_2$. For a spherical system $R_1 = R_2 = R$. From \ref{dndk} we can
write 
\begin{equation}
\rho_{i,MRE}(p,m_i,R)= 1 + {\frac { 6\pi^2} {pR}} f_{i,S} + {\frac {12\pi^2}
                         {({pR})^2}} f_{i,C} 
\end{equation}  
where the surface contribution to the density of states is
\begin{equation}
f_{i,S} = - {\frac {1} {8\pi}} (1 - {\frac {2} {\pi}}arctan {\frac {p} {m_i}}) 
\end{equation}
and the curvature contribution is given by Madsen's ansatz \cite {madsen}
\begin{equation}
f_{i,C} = {\frac {1} {12\pi^2}} [1 - {\frac {3p} {2m_i}} ({\frac {\pi} {2}} -
           arctan {\frac {p}{m_i}})]
\end{equation}
Using MRE formalism the density 
of states of the fireball can be reduced compared to the bulk one. For a range
of small momenta it becomes negative. This non-physical negative values are 
removed by introducing an infrared cutoff in momentum space \cite{kiriyama}.
We have to perform the following replacements in order to obtain the 
thermodynamic quantities
\begin{equation}
{\int_0^\propto} \ldots {\frac {d^3p} {{(2\pi)}^3}} \rightarrow 
{\int_{\lambda_{i,IR}}^\propto} \ldots \rho_{i,MRE} {\frac {d^3p} {{(2\pi)}^3}}
\end{equation}
The IR cut-off $\lambda_{i,IR}$ is the largest solution of the equation 
$\rho_{i,MRE}(p, m_i, R) = 0$ with respect to the momentum p. In 
\cite{deb1} the
finite volume effect was incorporated by introducing a lower momentum cutoff
$\lambda$ which depends on system size $R$ only. But here we have 
introduced a lower
momentum cutoff $\lambda_{IR}$ which is a function of mass. Thus for higher
masses the available phase space will be more restricted and will have lesser
contribution. Also in previous studies of finite system PNJL model, surface
and curvature effects were neglected for simplicity. In the next section
we will discuss the PNJL model considering the finite volume
effect through the infrared cut-off and surface and curvature effect
using the MRE technique. We have calculated the value of $\lambda_{IR}$ 
for different values of $R$ and $u$, $d$ and $s$ masses. Thus for $R= 2fm$, the lower cut-off for different masses are $\lambda_{u,IR} =75.2$
and $\lambda_{s,IR} = 131.0$ and for $R= 4fm$, the lower cut-offs are $\lambda_{u,IR}=40.0$ and $\lambda_{s,IR} = 75.1$. $u$ and $d$ quarks have same value for 
$\lambda_{IR}$.

\section{Thermodynamic Potential}

The thermodynamic potential for the multi-fermion interaction in MFA of the
PNJL model with MRE contribution can be written as,

\begin {eqnarray}
 \Omega &=& {\cal {U^\prime}}[\Phi,\bar \Phi,T]+2{g_S}{\sum_{f=u,d,s}}
            {{\sigma_f}^2}-{{g_D} \over 2}{\sigma_u}
          {\sigma_d}{\sigma_s}-6{\sum_f}{\int_{0}^{\Lambda}}
     {{d^3p}\over{(2\pi)}^3} \rho_{i,MRE} E_{pf}\Theta {(\Lambda-{ |\vec p|})}\nonumber \\
       &-&2{\sum_f}T{\int_0^\infty}{{d^3p}\over{(2\pi)}^3} \rho_{i,MRE}
       [\ln\left[1+3(\Phi+{\bar \Phi}e^{-{(E_{pf}-\mu)\over T}})
       e^{-{(E_{pf}-\mu)\over T}}+e^{-3{(E_{pf}-\mu)\over T}}\right]\nonumber\\
       & +& \ln\left[1+3({\bar \Phi}+{ \Phi}e^{-{(E_{pf}+\mu)\over T}})
            e^{-{(E_{pf}+\mu)\over T}}+e^{-3{(E_{pf}+\mu)\over T}}\right]]
\end {eqnarray}
where $E_{pf}=\sqrt {p^2+M^2_f}$ is the single quasi-particle energy,
$\sigma_f^2=(\sigma_u^2+\sigma_d^2+\sigma_s^2)$ and 
$\sigma_f^4=(\sigma_u^4+\sigma_d^4+\sigma_s^4)$.
In the above integrals, the vacuum integral
has a cutoff $\Lambda$ whereas the medium dependent integrals
have been extended to infinity. In the present study we have two sets of parameter sets
(a) PNJL-6-quark for $R=2 fm$, (b) PNJL-6-quark for $R=4 fm$.

\vskip 0.6in
{\subsection{Taylor expansion of pressure}}
The freeze-out curve $T(\mu_B)$ in the $T-\mu_B$ plane and the dependence of 
the baryon chemical potential on the center of mass energy in nucleus-nucleus 
collisions can be parametrized by \cite{cleymans} 
\begin {equation}
T(\mu_B) = a - b\mu_B^2 - c\mu_B^4
\end {equation}
where $a = (0.166 \pm 0.002) $ $GeV$, $b = (0.139 \pm 0.016) $ ${ GeV^{-1}}$,  
$c = (0.053 \pm 0.021) $ $GeV^{-3} $ and
\begin {equation}
\mu_B (\sqrt s_{NN}) = d/{(1+ e\sqrt s_{NN})}
\end {equation}
with $d$, $e$ given in {Table 1} in \cite{karsch-strange}.
The ratio of baryon to strangeness chemical potential on the freeze-out 
curve shows a weak dependence on the collision energy
\begin{equation}
{\mu_S\over\mu_B} \sim 0.164 + 0.018 \sqrt s_{NN}
\end{equation}
  
The pressure of the strongly interacting matter can be written as,
\begin {equation}
P(T,\mu_B,\mu_Q,\mu_S)=-\Omega (T,\mu_B,\mu_Q,\mu_S),
\label{pres}
\end {equation}
where $T$ is the temperature, $\mu_B$ is the baryon (B) chemical potential, 
$\mu_Q$ is the charge (Q) chemical potential and $\mu_S$ is the 
strangeness (S) chemical potential. From the usual thermodynamic
relations the first derivative of pressure with respect to
quark chemical potential $\mu_q$ is the quark number density and
the second derivative corresponds to the quark number susceptibility (QNS).

 Minimizing the thermodynamic potential numerically with
respect to the fields $\sigma_u$, $\sigma_d$, $\sigma_s$, $\Phi$ and 
$\bar \Phi$, the mean field value for pressure can be obtained 
using the equation (\ref{pres}) \cite {deb}.
The scaled pressure obtained in a given range of chemical potential 
at a particular temperature can be expressed in a Taylor series as,
\begin {equation}
\frac{p(T,\mu_B,\mu_Q,\mu_S)}{T^4}=\sum_{n=i+j+k}c_{i,j,k}^{B,Q,S}(T) 
           (\frac{\mu_B}{T})^i (\frac{\mu_Q}{T})^j (\frac{\mu_S}{ T})^k
\end{equation}
where,
\begin {equation}
c_{i,j,k}^{B,Q,S}(T)={\frac{1}{i! j! k!} 
\frac{\partial^i}{\partial (\frac{\mu_B}{T})^i} 
\frac{\partial^j}{\partial (\frac{\mu_Q}{T})^j} 
\frac{\partial^k {(P/T^4)}}{\partial (\frac{\mu_S}{T})^k}}\Big|_{\mu_{q,Q,S}=0}
\end{equation}
where $\mu_B$, $\mu_Q$, $\mu_S$ are related to the flavor chemical potentials  
$\mu_u$, $\mu_d$, $\mu_s$ as,  
\begin {equation}
  \mu_u=\frac{1}{3}\mu_B+\frac{2}{3}\mu_Q,~~~ 
  \mu_d=\frac{1}{3}\mu_B-\frac{1}{3}\mu_Q,~~~
  \mu_s=\frac{1}{3}\mu_B-\frac{1}{3}\mu_Q-\mu_S
\label{mureln1}
\end {equation}
In this work we evaluate the correlation coefficients up to fourth order
which are generically given by;
\begin{equation}
c_{i,j}^{X,Y}=\dfrac{1}{i! j!}\dfrac{\partial^{i+j}\left(P/T^4\right)}
{{\partial\left({\frac{\mu_X}{T}}\right)^i}{\partial\left({\frac{\mu_Y}{T}}
\right)^j}}
\end{equation}
where, X and Y each stands for B, Q and S with $X\neq Y$.
To extract the Taylor coefficients, first the pressure is obtained as 
a function of different combinations of chemical potentials for each 
value of T and fitted to a polynomial about zero chemical potential
using the gnu-plot fit program \cite{gnu}. Stability of the fit has 
been checked by varying the ranges of fit and simultaneously keeping 
the values of least squares to $10^{-10}$ or even less. At low temperature
fluctuations of a particular charge are dominated by lightest hadrons carrying
that charge. The dominant contribution to $\chi^2_B$ at low temperatures 
comes from protons (lightest baryon), while $\chi^2_S$ receives leading 
contribution from kaons (lightest strange hadron) and $\chi^2_Q$ from 
pions (lightest charged hadron). Since, pion is lighter than proton and kaon,
magnitude of $\chi^2_Q$ is more than that of $\chi^2_B$ and $\chi^2_S$.

\vskip 0.2in
{\subsection{Results}}
The experimental results for volume independent cumulant ratios of net-proton, 
net-kaon and net-charge number distribution are presented for all BES energies ${\sqrt s_{NN}}= 7.7, 11.5, 14.5, 19.6, 27, 39, 62.4$ and $200 \text{GeV}$ for top central and peripheral collisions {\cite{STAR-proton, STAR-charge, STAR-kaon}}. We have presented our results for finite size PNJL model including the MRE formalism for six-quark interactions to compare our results with experiment. The ratios of color charge fluctuations for different moments have been considered as they are independent of definitions of the interaction volume and also are more sensitive to produce correlation length.  

\begin{figure}[htb]
\centering 
 \includegraphics[scale=0.6]{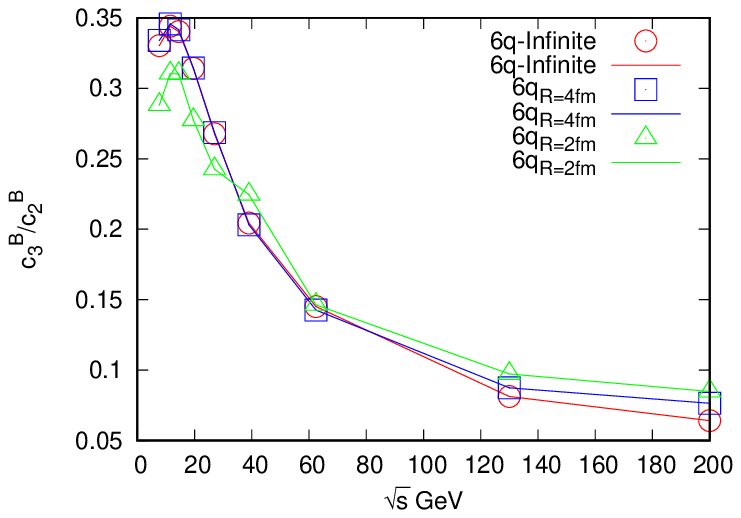}
 \includegraphics[scale=0.6]{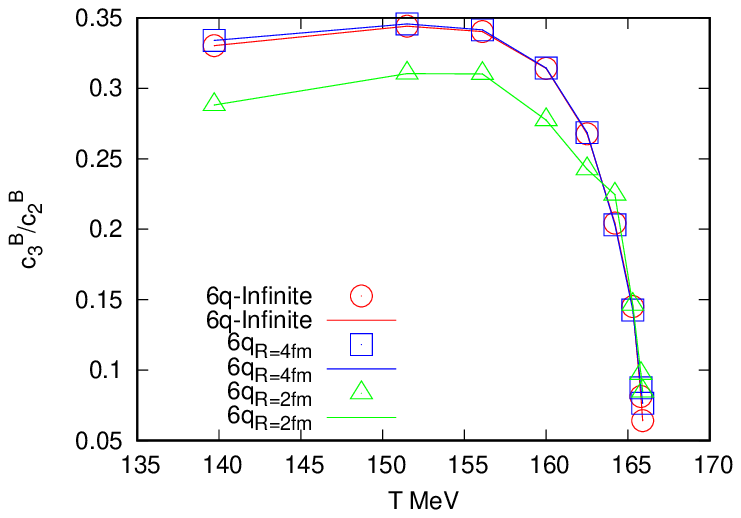}
 \includegraphics[scale=0.6]{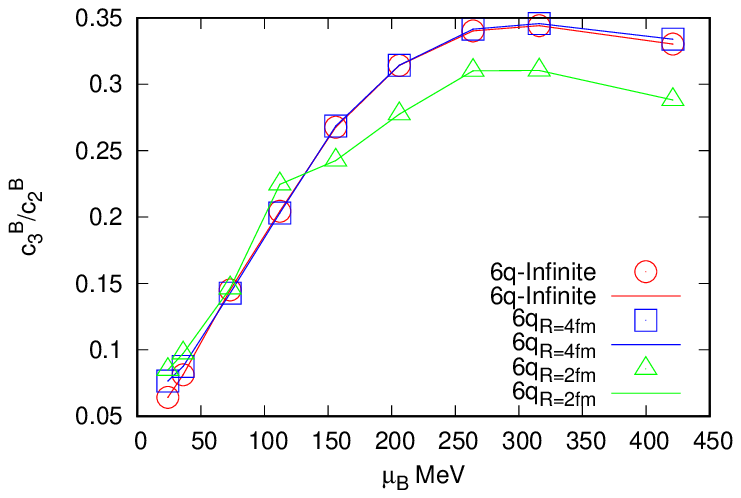}
 \caption{(Color online) Skewness for baryon fluctuation have been plotted with
 respect to collision centrality, baryon chemical potential and temperature. 
 PNJL model with infinite volume has been denoted by red color, PNJL model with $R=4 fm$ has been denoted by blue color and PNJL model with $R=2 fm$ has been
 denoted by green color.}
\label{c3c2B}
\end{figure}
The correlation length and the magnitude of fluctuations are important quantity as they diverge
at the critical end point. Because of the finite size and time effects, the correlation length takes a finite volume in 
the range $\sim 2-3 fm$. Higher non-Gaussian moments such as skewness 
and kurtosis can provide much better handle in location of CEP as they are much more sensitive than variance to the correlation length. So the moment products have been constructed to cancel out volume dependence. In this figure panel \ref{c3c2B}, we have plotted the skewness of
baryon fluctuation with respect to collision energy, temperature and baryon 
chemical potential for the PNJL model with infinite volume and also for the 
finite size with $R=2fm$ and $R=4fm$. The plots show a sharp rise near $\sqrt{s}= 130 GeV$, $\mu_{B}= 36 MeV$ and $T=165.8 MeV$ for PNJL model with $R=2fm$ compared to other sets of the model. The sharp rise may be the indication of critical region. In figure \ref{c4c2B} we have plotted the kurtosis for baryon fluctuation with respect to collision centrality, temperature and baryon chemical potentials. The
kurtosis shows a minimum around $20-40 GeV$ for collision centrality and a maximum
around $40 MeV$ for baryon chemical potential. Also it shows a sharp rise in temperature at $165 MeV$. Both skewness and kurtosis of baryon fluctuations are in
good agreement with the experimental result qualitatively \cite{STAR-proton}. 

\begin{figure}[htb]
\centering 
 \includegraphics[scale=0.6]{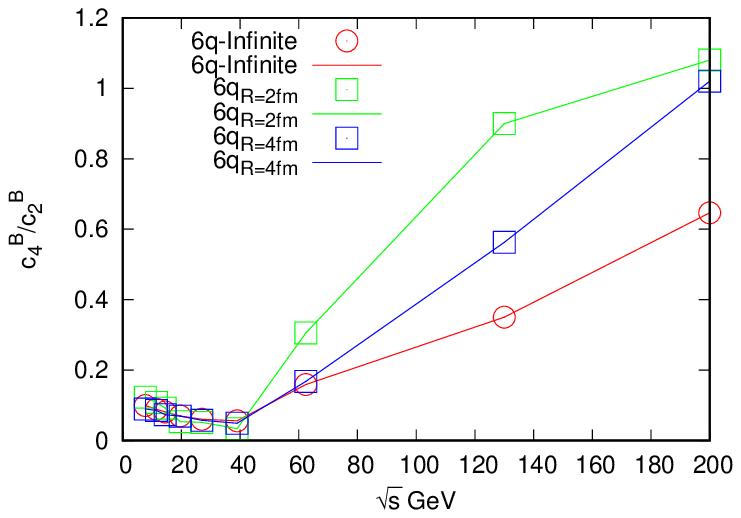}
 \includegraphics[scale=0.6]{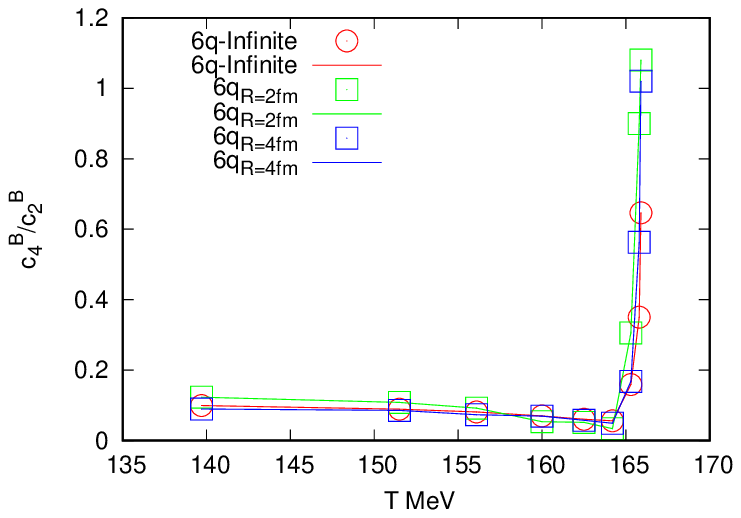}
 \includegraphics[scale=0.6]{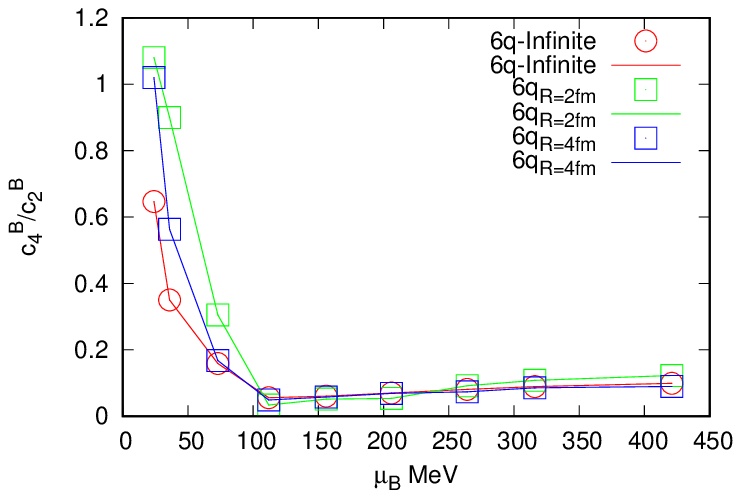}
 \caption{(Color online) Kurtosis for baryon fluctuation have been plotted with
  respect to collision centrality, baryon chemical potential and temperature. 
  PNJL model with infinite volume has been denoted by red color, PNJL model with $R=4 fm$ has been denoted by blue color and PNJL model with $R=2 fm$ has been denoted by green color. }
\label{c4c2B}
\end{figure}

\begin{figure}[htb]
\centering 
 \includegraphics[scale=0.6]{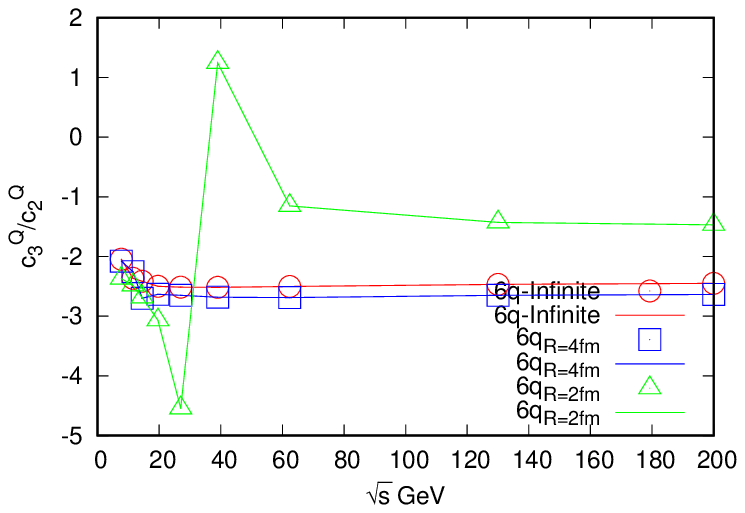}
 \includegraphics[scale=0.6]{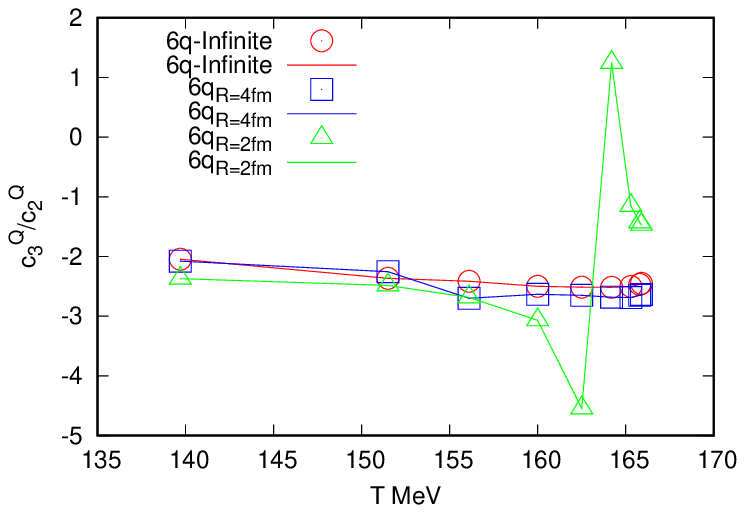}
 \includegraphics[scale=0.6]{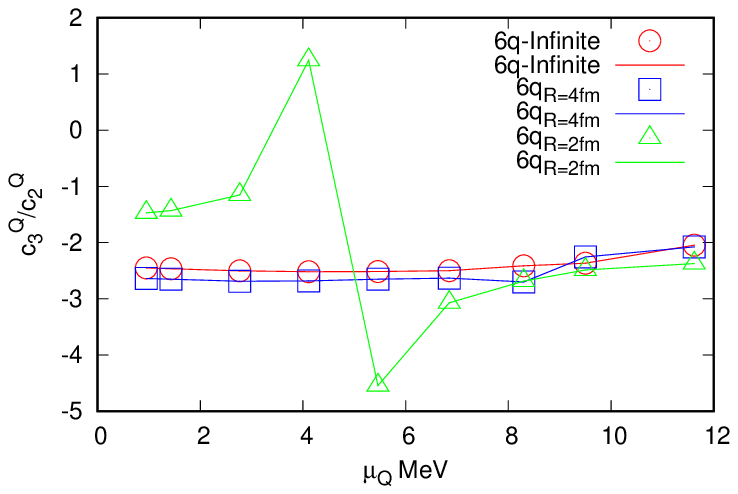}
 \caption{(Color online) Skewness for charge fluctuation have been plotted with
  respect to collision centrality, charge chemical potential and temperature. 
  PNJL model with infinite volume has been denoted by red color, PNJL model with $R=4 fm$ has been denoted by blue color and PNJL model with $R=2 fm$ has been denoted by green color.}
 \label{c3c2Q}
\end{figure}
Figure \ref{c3c2Q} shows the skewness for charge fluctuation with respect to collision centrality, charge chemical potential and temperature. For $R=2 fm$,
the plots show a sharp rise around collision energy $20-40 GeV$, temperature around $165 MeV$ and chemical potential around $\mu_Q= 4 MeV$. The value of fluctuation is significantly large for $R=2 fm$ than for $R=4 fm$ and infinite 
volume. Figure \ref{c4c2Q} represents the kurtosis of charge fluctuation with respect to 
collision centrality, charge chemical potential and temperature. The kurtosis 
shows a minimum around $\sqrt{s}=20-40 GeV$ for both finite and infinite volume. 
For temperature and charge chemical potential there is a sharp increase in fluctuation for both finite and infinte volume system. However the peaks are not
in same position. It shifts to higher charge chemical potential and lower temperature for infinite volume system. For $R=2fm$ the peak is around 
$\mu_Q=5.5 MeV$ and $T=163 MeV$ and for infinite volume the peak is around 
$\mu_Q=8 MeV$ and $T=155 MeV$. Both the results for skewness and kurtosis are
qualitatively similar to the experimental data \cite{STAR-charge} .

\begin{figure}[htb]
\centering 
 \includegraphics[scale=0.6]{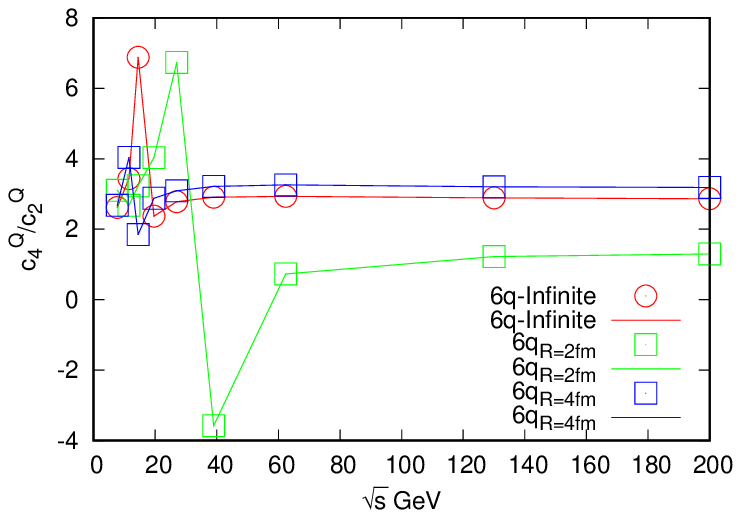}
 \includegraphics[scale=0.6]{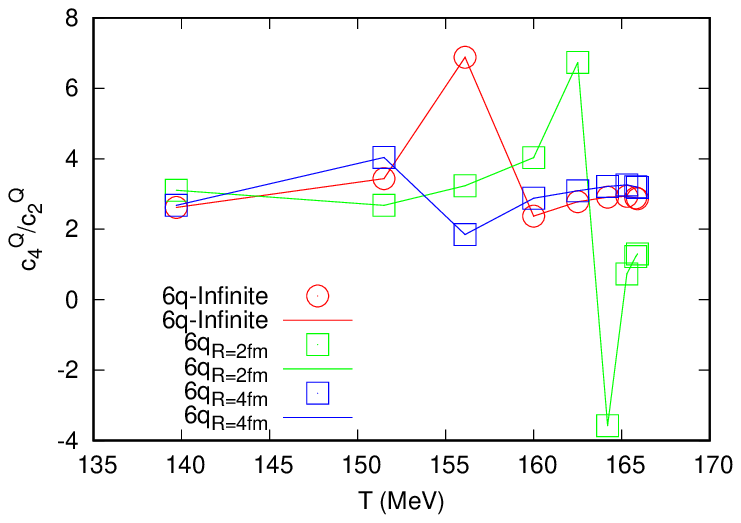}
 \includegraphics[scale=0.6]{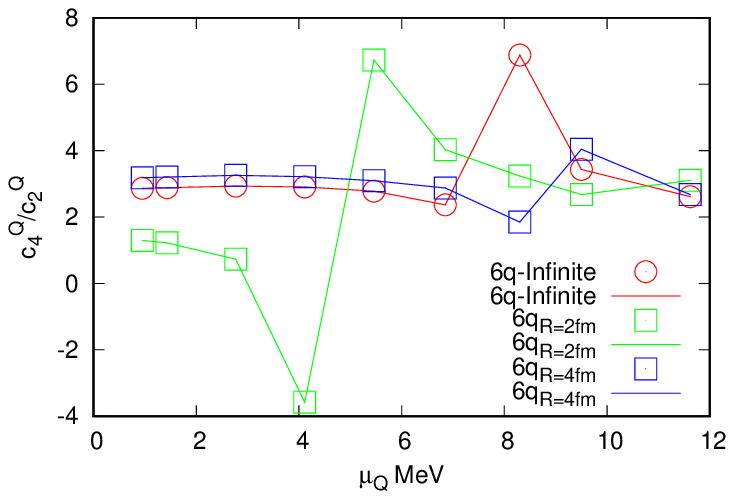}
 \caption{(Color online) Kurtosis for charge fluctuation have been plotted with
  respect to collision centrality, charge chemical potential and temperature. 
  PNJL model with infinite volume has been denoted by red color, PNJL model with $R=4 fm$ has been denoted by blue color and PNJL model with $R=2 fm$ has been
  denoted by green color. }
 \label{c4c2Q}
\end{figure}

\begin{figure}[htb]
\centering 
 \includegraphics[scale=0.6]{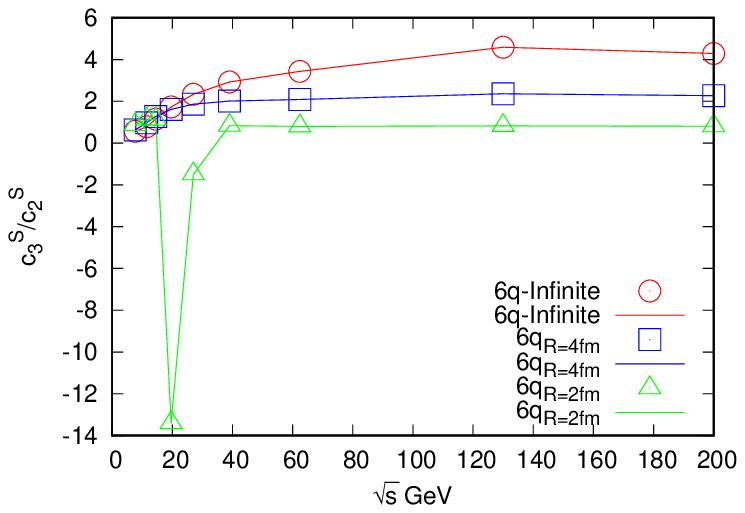}
 \includegraphics[scale=0.6]{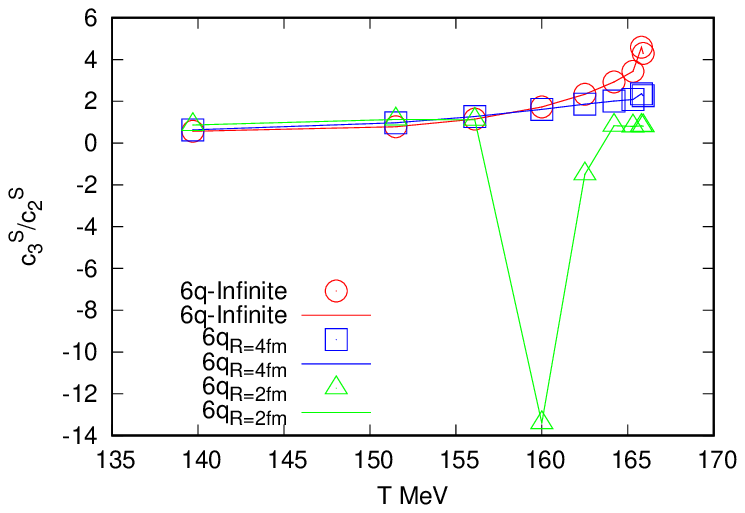}
 \includegraphics[scale=0.6]{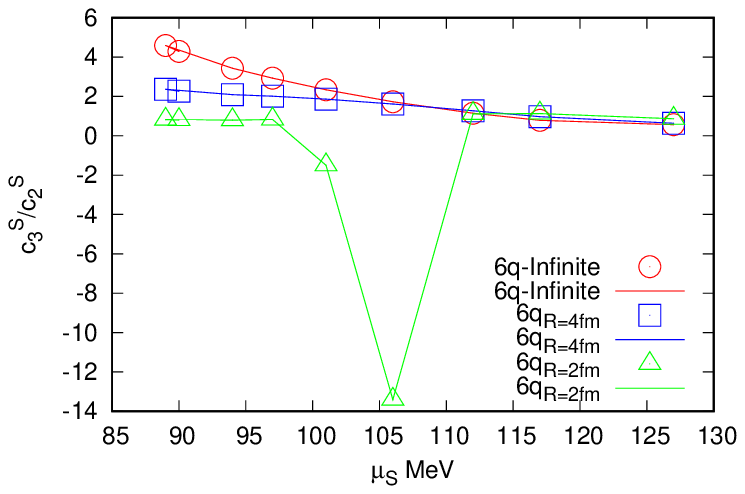}
 \caption{(Color online) Skewness for strangeness fluctuation have been plotted with respect to collision centrality, strangeness chemical potential and temperature. PNJL model with infinite volume has been denoted by red color, PNJL model with $R=4 fm$ has been denoted by blue color and PNJL model with $R=2 fm$ has been denoted by green color.}
 \label{c3c2S}
\end{figure}

Figure ({\ref {c3c2S}}) and ({\ref{c4c2S}}) shows the variation of 
$C_3 / C_{2_S}$ and $C_4 / C_{2_S}$ fluctuations with respect to 
different collision energies, strangeness chemical potential and temperature for PNJL model. Skewness shows more fluctuation near $\sqrt{s}=20 GeV$, 
$\mu_S=105 MeV$ and $T=160 MeV$. The fluctuation is more prominent for $R=2fm$. Kurtosis also shows similar kind of behaviour as skewness. In recent experimental data no significant deviation has been found with respect to the Poisson expectation value within statistical and systematic
uncertainties for both the moments {\cite {STAR-kaon}}. The results from the PNJL model are in good agreement
with the experimental results.  
For the collision energy $ \sqrt s <  27 \text{ GeV}$, there is 
an enhancement of fluctuation for PNJL model. Also in case of STAR results, 
there is an deviation from Poisson expectation value.  
Although the results for both skewness and kurtosis have 
qualitative similarities for 
PNJL model, the values have quantitative differences. 

\begin{figure}[htb]
\centering 
 \includegraphics[scale=0.6]{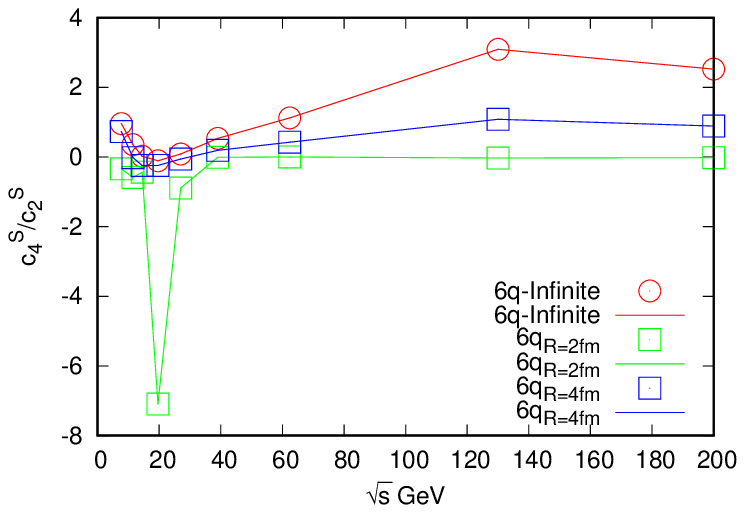}
 \includegraphics[scale=0.6]{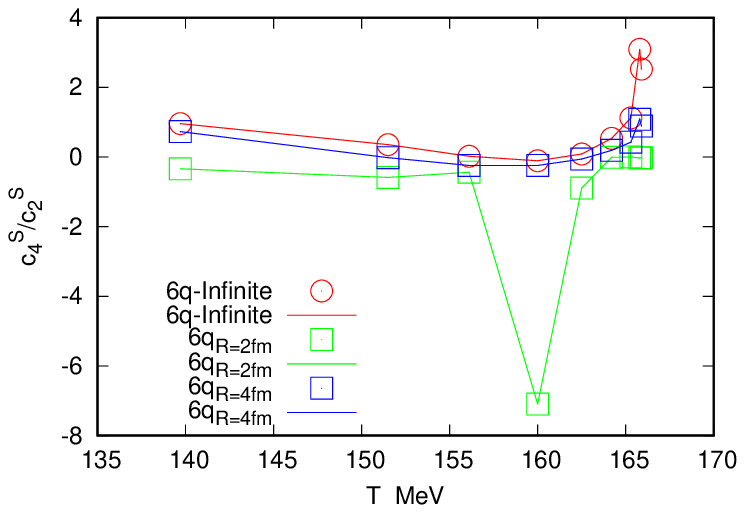}
 \includegraphics[scale=0.6]{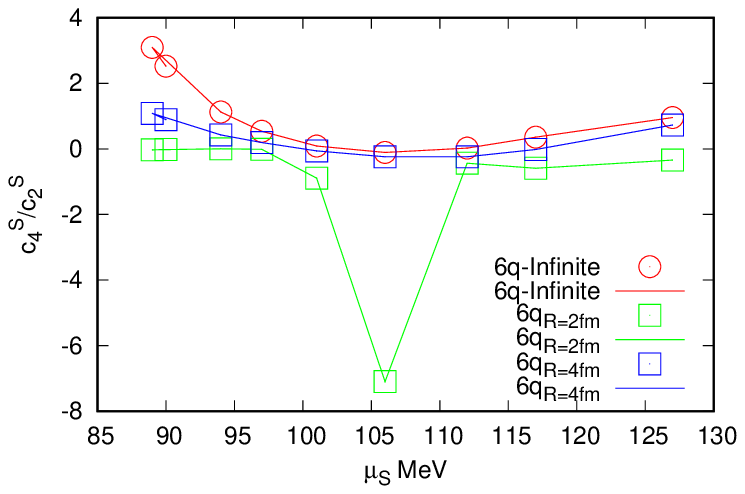}
 \caption{(Color online) Kurtosis for strangeness fluctuation have been plotted with respect to collision centrality, strangeness chemical potential and temperature. PNJL model with infinite volume has been denoted by red color, PNJL model with $R=4 fm$ has been denoted by blue color and PNJL model with $R=2 fm$ has been
  denoted by green color. }
 \label{c4c2S}
\end{figure}

\section{summary}
We have discussed properties of net baryon, charge and strangeness fluctuations in nuclear matter within finite size PNJL model. The ratio of fourth to second order moment and the third to the second order moment of fluctuations are considered as observables. All the correlations were obtained by fitting the pressure in a Taylor series expansion around the finite baryon, charge and strangeness chemical potentials. These chemical potentials are obtained from the freeze-out curve which depends on the collision energies in the BES scan at the heavy-ion collision experiment. The results are shown for PNJL model with 6 quark interactions and two finite volume systems with lateral size $R=2 fm$ and $R=4fm$. We have considered the
multiple reflection expansion method to include the surface and curvature
effect of the fire ball along with the volume effect.

Skewness and kurtosis of baryon, charge and strangeness fluctuations in PNJL model have similar features along the collision energy of heavy ion experiments. As we increase the temperature both skewness and kurtosis value decreases quantitatively. For collision energy less than $27$ $GeV$, the value of kurtosis and skewness are higher. The recent experimental observations show small deviations for skewness and kurtosis for low collision energy. Similarly in PNJL model we have found an enhancement of fluctuations for low collision energy less than $27 GeV$. Also, near the transition temperature the skewness ratio is very near to the Poisson expectation value.         

The study of various equilibrium thermodynamic measurements of the correlators using PNJL model would be helpful in determining the finite temperature finite density behavior of the hadronic sector. comparison of PNJL results with the experimental value will ensure the understanding of the physics behind the critical region and to locate the critical point in the strongly interacting matter.   

\acknowledgements P.D would like to thank Women Scientist Scheme A (WOS-A) of Department of Science and Technology (DST) funding with grant no SR/WOS-A/PM-10/2019 (GEN).

\end{document}